\documentclass[aps,prl,floatfix,twocolumn,superscriptaddress]{revtex4-1}
\usepackage{amsmath,amssymb}
\usepackage{graphicx}
\usepackage{epsfig}
\usepackage{ulem}
\usepackage[usenames]{color}

\begin{document}

\title{Proposal for Plasmon Spectroscopy of Fluctuations\\ in Low-Dimensional Superconductors}






\author{V.~M.~Kovalev}
\affiliation{A.V.~Rzhanov Institute of Semiconductor Physics, Siberian Branch of Russian Academy of Sciences, Novosibirsk 630090, Russia}
\affiliation{Novosibirsk State Technical University, Novosibirsk 630073, Russia}

\author{I.~G.~Savenko}
\affiliation{Center for Theoretical Physics of Complex Systems, Institute for Basic Science (IBS), Daejeon 34126, Korea}
\affiliation{Basic Science Program, Korea University of Science and Technology (UST), Daejeon 34113, Korea}

\date{\today}

\begin{abstract}
We propose to unleash an optical spectroscopy technique to monitor the superconductivity and properties of superconductors in the fluctuating regime.
This technique is operational close to the plasmon resonance frequency of the material, and it intimately connects with the superconducting fluctuations slightly above the critical temperature $T_c$.
%
We find the Aslamazov-Larkin corrections to AC linear and DC nonlinear electric currents in a generic two-dimensional system exposed to an external longitudinal electromagnetic field.
%
First, we study the plasmon resonance of normal electrons  near $T_c$, taking into account their interaction with superconducting fluctuations, and show that fluctuating Cooper pairs reveal a redshift of the plasmon dispersion and an additional mechanism of plasmon scattering, which surpasses both the electron-impurity  and the Landau dampings.
Second, we demonstrate the emergence of a drag effect of superconducting fluctuations by the external field resulting in considerable, experimentally measurable corrections to the electric current in the vicinity of the plasmon resonance.
\end{abstract}

\maketitle


\textit{Introduction.---}The study of fluctuating phenomena in superconductors is a wide field of modern research~\cite{Narlikar, Ketterson, LarkinVarlamov}.
At the temperature approaching 
$T_c$ from above, there start to emerge (and collapse) Cooper pairs even before the system reaches $T_c$.
It results in fluctuations of the Cooper pairs density, which might sufficiently modify the conductivity of the system.
This effect is especially pronounced in samples of reduced dimensionality, as Aslamazov and Larkin (AL) reported in their pioneering work~\cite{AL}.
Later their theory was developed further to study high-frequency phenomena in superconductors in the \textit{fluctuating regime}~\cite{VBLM, AV}
and the fluctuating corrections in linear transport phenomena in superconductors, such as the Hall effect~\cite{LarkinVarlamov}, thermoelectric phenomena~\cite{thermo}, and the critical viscosity of electron gas~\cite{Galitskii}.
\textcolor{black}{In the mean time, superconducting optoelectronics is becoming a rapidly growing field of modern research~\cite{RefAsano, RefSasakura, RefGodschalk, OurAV, OurKus2}.}

In this Letter, we demonstrate that it is possible to monitor and manipulate transport of carriers of charge in superconductors using external electromagnetic (EM) waves \textcolor{black}{with plasmonic frequencies, 
interacting with the
superconducting fluctuations (SFs) due to their coupling with normal electrons.}
We develop a theory of linear AC and second-order DC response of a two-dimensional (2D) electron gas (2DEG) in the vicinity of the plasmon resonance and $T_c$, where the SFs play an essential role.
As a first step, we study the plasmon oscillations of normal electrons in the presence of the gas of fluctuating Copper pairs.
Second, we find the fluctuating corrections to the drag effect, which consists in the emergence of a stationary electric current as the second-order response to an external alternating EM perturbation of the system~\cite{GlazovReview}.
It should be noted, that while exciting plasmons, the internal induced long-range Coulomb fields activate.
They act on both the 
electrons and the fluctuating Cooper pairs.
In other words, the interaction between electrons and SFs cannot be disregarded, as it is usually done when considering static and dynamic corrections to the Drude conductivity due to the presence of an external uniform EM field in superconductors above $T_c$. Such an interaction strongly modifies the plasmon modes of normal electrons and opens a new microscopic mechanism of their damping and a spectroscopy tool to study SFs.

The general approach to the description of fluctuations in superconductors above 
$T_c$ relies on rather cumbersome methods of quantum field theory, \textcolor{black}{or phenomenological Ginsburg-Landau theory~\cite{LarkinVarlamov}}.
However, as it was first pointed out by AL, 
it is often sufficient to use a considerably simpler approach based on the Boltzmann kinetic equations, which disregards the wave nature of fluctuating Cooper pairs and operates with the quasiparticle picture~\cite{CommentParaC}.
%
%
%
%
We will use the Boltzmann equations to calculate the AL corrections in the response of a 2DEG to an external longitudinal EM field $\textbf {E} (\textbf {r}, t) = (E (\textbf {r}, t), 0)$, with $ E (\textbf {r}, t) = E_0 \cos {(ikx-i \omega t)} $, which directs along the plane of the quantum well ($xy$ plane), containing the electron gas.
Such a setup arises (i) when studying acoustoelectric effects in two-dimensional systems~\cite{ParmenterAEE, RefWixforth, RefWillet, graphene1, OurPRLVAE}, 
(ii) in photo-induced transport in two-dimensional systems (e.g., the photon drag effect) \cite{WSP, EM}, (iii) when plasma waves are excited \cite{Plasmonics}, and also (iv) in ratchet effects in two-dimensional systems \cite{Ivchenko1, Ivchenko2, Ivchenko3, Ivchenko4, Ivchenko5}. 
In particular, it has recently been shown, that a photoinduced ratchet current can be sufficiently enhanced in the vicinity of the plasmon resonance~\cite{Kachorovskii}.
This finding and the details of the approach used in Ref.~\onlinecite{Kachorovskii} let us hypothesize that there might be many phenomena, which become enhanced near the plasmon resonance.

%

%
%
%
\begin{figure}
\includegraphics[width=.48\textwidth]{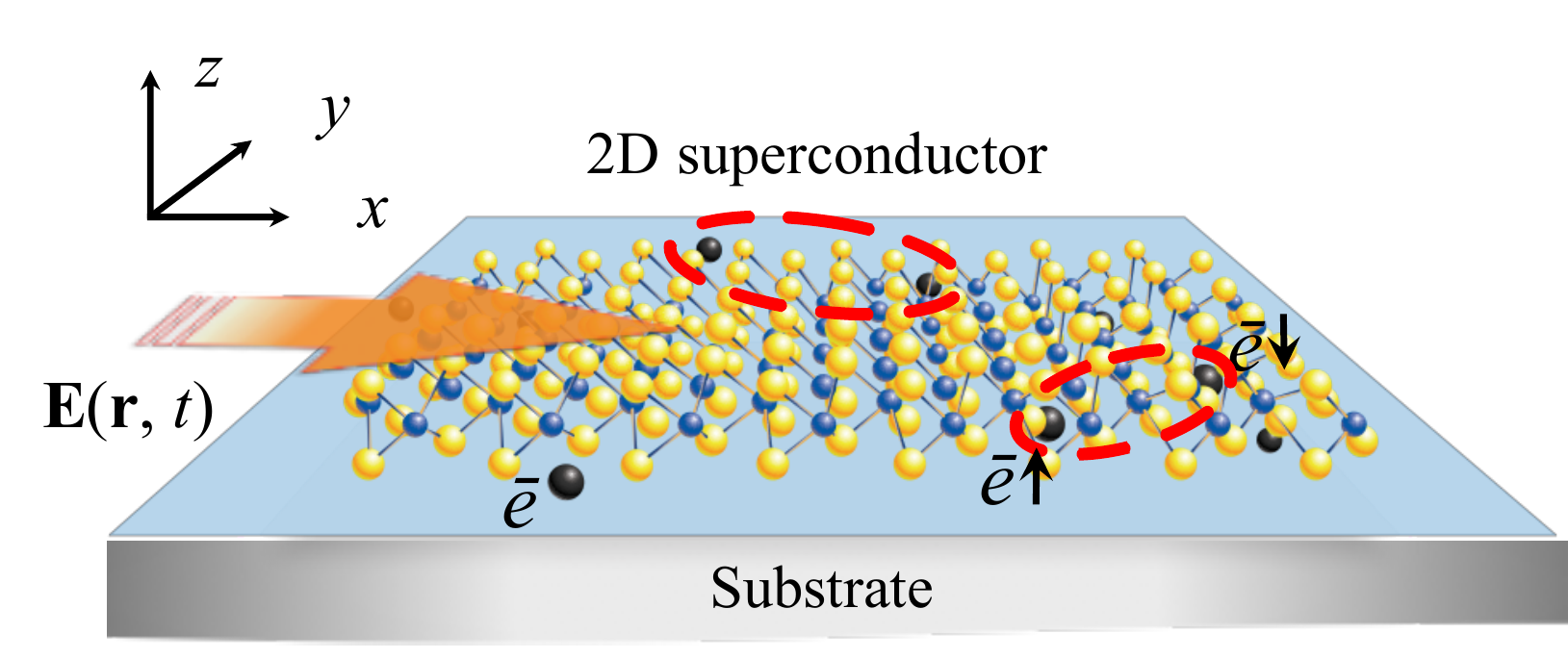}
\caption{System schematic. A two-dimensional material on a substrate at the temperature close to $T_c$. 
The system is exposed to a longitudinal EM field $\mathbf{E}$.}
\label{Fig1}
\end{figure}
%
%
%


{\textit{Plasmon resonance of 2DEG in the presence of SFs.---}}
Following the standard approach~\cite{vitlinachaplik, sarma, chapliknew}, we consider
%
the wave vector $\mathbf{k}$ and frequency $\omega$-dependent
dielectric function of the 2DEG $\varepsilon(\textbf{k},\omega)$, 
taking into account the  SFs~\cite{CommentBoseMixture}.
In the absence of external perturbations, the Cooper pairs obey the classical Rayleigh-Jeans distribution $f_0(\mathbf{p})=T/\varepsilon_\textbf{p}$, where $\mathbf{p}$ is a center-of-mass momentum of the Cooper pair, the temperature $T$ is taken in energy units, and $\varepsilon_\textbf{p}=\alpha T_c(\epsilon+\xi^2p^2/\hbar^2)=p^2/4m+\alpha T_c\epsilon$ is the energy with $\hbar$ the Planck's constant and $\epsilon=(T-T_c)/T_c>0$ the reduced temperature~\cite{LarkinVarlamov}; $\alpha$ is fixed by the relation 
$4m\alpha T_c\xi^2/\hbar^2=1$, where $m$ is an electron effective mass; 
the coherence length $\xi$ in 2D samples has different definitions for the cases of clean $T\tau/\hbar\gg1$ and disordered $T\tau/\hbar\ll1$ regimes, where $\tau$ is electron relaxation time (which we assume constant for simplicity).
Both the regimes are sewn in the general expression
\begin{gather}\label{Eqq1}
\xi^2=\frac{v_F^2\tau^2}{2}\Bigl[
\psi\left(\frac{1}{2}\right)
-
\psi\left(\frac{1}{2}+\frac{\hbar}{4\pi T\tau}\right)
+\frac{\hbar\psi'\left(\frac{1}{2}\right)}{4\pi T\tau}\Bigr],
\end{gather}
where $\psi(x)$ is the digamma function and $v_F=\hbar \sqrt{4\pi n}/m$ is the Fermi velocity.
%
%
%
%
%
%

The internal induced electric field $\textbf{E}^{i}(k,\omega)$ due to the fluctuations of the charge 
densities can be found from the Poisson equation in the quasistatic limit, when we can neglect the retardation effects.
Assuming that $z$ axis is directed across the 2D system, which is located on a substrate ($z<0$) with a dielectric constant $\kappa$ (Fig.~\ref{Fig1}), and using the ansatz $\exp(ikx-i\omega t)$ for all the time and position-dependent quantities, we find the Poisson equation for the scalar potential $\varphi(z)$ of the induced field in the form~\cite{CommentGauss}
\begin{gather}\label{Eqq2}
\left(\frac{\partial}{\partial z}\kappa(z)\frac{\partial}{\partial z}-k^2\right)\varphi(z)=-4\pi(\rho_{k\omega}+\varrho_{k\omega})\delta(z),
\end{gather}
where $\kappa(z)=1$ for $z>0$ and $\kappa(z)=\kappa$ for $z<0$; $\rho_{k\omega}$ and $\varrho_{k\omega}$ are Fourier-transforms of charge densities due to the normal electrons and fluctuating Cooper pairs, respectively. 
Solving Eq.~\eqref{Eqq2}, we find
%
\begin{gather}\label{Eqq3}
\varphi(z)=\frac{4\pi}{(\kappa+1)k}e^{-k|z|}\left(\rho_{k\omega}+\varrho_{k\omega}\right).
\end{gather}

Furthermore, using the continuity equation for both the components of the charge density and expressing the currents via conductivities, we come to the system of equations
\begin{gather}\label{Eqq4}
\rho_{k\omega}=-i\frac{k^2\sigma^D_{k\omega}}{\omega}\varphi(0),\\\nonumber
\varrho_{k\omega}=-i\frac{k^2\sigma^{AL}_{k\omega}}{\omega}\varphi(0),
\end{gather}
where $\sigma^D_{k\omega}$ and $\sigma^{AL}_{k\omega}$ are Drude and
Aslamazov-Larkin conductivities. 
The determinant of the system~(\ref{Eqq4}),
\begin{gather}
\label{Eqq5}
\varepsilon(\textbf{k},\omega)=1+i\frac{4\pi k}{(\kappa+1)\omega}\left(\sigma^D_{k\omega}+\sigma^{AL}_{k\omega}\right),
\end{gather}
\textcolor{black}{allows us to find the dispersion relation of collective modes and their damping by putting $\varepsilon(\textbf{k},\omega)=0$.} 
The plasmon pole lies in the frequency range $\omega\gg kv_F$. Since $v_F\gg u$, where $u=p/2m$ is the Cooper pair velocity, we can disregard the spatial dispersions of both the conductivities, yielding
%
\begin{gather}
\label{Eqq6}
\sigma^{D}_{\omega}=e^2\int\frac{d\textbf{p}}{(2\pi\hbar)^2}\frac{v_x^2\tau}{1-i\omega\tau}\left(-\frac{\partial \mathcal{F}_0}{\partial\tilde\varepsilon_\textbf{p}}\right),\\
\label{Eqq62}
\sigma^{AL}_{\omega}=(2e)^2\int\frac{d\textbf{p}}{(2\pi\hbar)^2}\frac{u_x^2\tau_\textbf{p}}{1-i\omega\tau_\textbf{p}}\left(-\frac{\partial f_0}{\partial\varepsilon_\textbf{p}}\right),
\end{gather}
where $v_x$, $\tilde\varepsilon_\textbf{p}=p^2/2m$, and $\mathcal{F}_0$ are the velocity, energy, and equilibrium Fermi distribution function of normal electrons, and $\tau_\textbf{p}=\hbar\pi\alpha/(16\varepsilon_\textbf{p})$ is the Cooper pair lifetime. 

Using~\eqref{Eqq6}, 
we  rewrite Eq.~\eqref{Eqq5} in the form
\begin{gather}\label{Eq7}
\left(\frac{\omega}{\omega_p}\right)^2+
i\left[\frac{1}{\omega_p\tau}+\omega_p\tau\frac{\sigma^{AL}_{\omega}}{\sigma^D_0}\right]\left(\frac{\omega}{\omega_p}\right)
-\frac{\sigma^{AL}_{\omega}}{\sigma^D_0}
-1=0,
\end{gather}
%
where $\omega_p^2=4\pi e^2nk/m(\kappa +1)$ \textcolor{black}{ is a bare plasmon frequency for 2D electron gas} and $\sigma^D_0=e^2n\tau/m$ \textcolor{black}{ is a static Drude conductivity.} 
Furthermore,  \textcolor{black}{introducing a dimensionless variable} $x=\varepsilon_\textbf{p}/\alpha T_c\epsilon$ in~\eqref{Eqq62}, we rewrite
\begin{gather}\label{Eq8}
\sigma^{AL}_{\omega}=\sigma^{AL}_0\int\limits_1^\infty\frac{dx}{x^2}\frac{2(x-1)}{x-i\beta_\omega},
\end{gather}
where $\sigma^{AL}_0=e^2/(16\hbar\epsilon)$ is a static AL conductivity and  $\beta_\omega=\pi\hbar\omega/(16T_c\epsilon)$ contains all the frequency dependence. 
A typical range of plasmon frequencies is $\omega_p\sim10^{10}\div10^{11}$~s$^{-1}$~\cite{Kukushkin}, and for $T_c=10$~K and $\epsilon=0.1$ we find $\beta_{\omega_p}\sim0.01\div0.2$. 
It means that the  electromagnetic field induced by the plasmon oscillations of normal electrons is quasi-static for the fluctuating Cooper pairs, and we can safely disregard the frequency dependence of AL conductivity in the vicinity of the plasmon resonance. 
Then Eq.~(\ref{Eq7}) has an exact solution~\cite{CommentRegPlasmon},
\begin{gather}\label{Eq9}
\omega=\omega_p\sqrt{1+\frac{\sigma^{AL}_0}{\sigma^{D}_0}-\left(\frac{1}{2\omega_p\tau}+\frac{\omega_p\tau}{2}\frac{\sigma^{AL}_0}{\sigma^{D}_0}\right)^2}\\
\nonumber
-\frac{i}{2}\left(\frac{1}{\tau}+\omega_p^2\tau\frac{\sigma^{AL}_0}{\sigma^{D}_0}\right).
\end{gather}
%
%
%
%
%
%
%
%
%
Assuming $\sigma^{AL}_0\ll\sigma^{D}_0$ 
and $\omega_p\tau\gg1$, we find~\cite{CommentBigBeta}
\begin{gather}\label{Eq11}
\omega=\omega_p\sqrt{1-\left(\frac{\omega_p\tau}{2}\frac{\sigma^{AL}_0}{\sigma^{D}_0}\right)^2}
-i\frac{\omega_p^2\tau}{2}\frac{\sigma^{AL}_0}{\sigma^{D}_0}.
\end{gather}
Relation~\eqref{Eq11} represents the first central result of this Letter.
We immediately see, that \textcolor{black}{even if we take a small factor 
$\sigma^{AL}_0/\sigma^{D}_0\ll1$, it can be compensated by the large (plasmonic) factor $\omega_c\tau\gg 1$, making their product arbitrary~\cite{CommentLevanyuk}.} 
It means that the interaction of normal electrons with fluctuating Cooper pairs leads to a significant renormalization of both the plasmon dispersion (redshift)
and its damping.

\textcolor{black}{
The plasmon branch exists when the expression under the square root in~\eqref{Eq11} is positive,
\begin{gather}\label{Eq12}
\eta=\frac{\omega_p\tau}{2}\frac{\sigma^{AL}_0}{\sigma^{D}_0}<1.
\end{gather}
Moreover, the \textcolor{black}{absolute value of} the damping \textcolor{black}{ $\Gamma_{s}=|\textmd{Im}\,\omega|$} should be smaller than \textcolor{black}{$\textmd{Re}\,\omega$. 
In other words,} $\eta/\sqrt{1-\eta^2}<1$ or $\eta<\sqrt{2}/2$; 
then plasmons represent ``good'' quasiparticles~\cite{RefBruus}. 
For example, if $\eta=0.6$, the relative shift of the plasmon frequency $\delta\omega_p/\omega_p=20~\%$, which is fully detectable experimentally.}
%
%
%
%
%
%

Let us compare different plasmon damping mechanisms. 
One of them is due to the scattering of normal electrons with impurities,  $\Gamma_i=1/2\tau$~\cite{CommentRegPlasmon}.
The ratio of imaginary part of Eq.~(\ref{Eq11}) and $\Gamma_i$ is $\Gamma_s/\Gamma_i=(\omega_p\tau)\eta$. Despite $\eta<1$, 
the fluctuations-induced plasmon damping can exceed the impurity-induced one (since $\omega_p\tau\gg1$)~\cite{CommentWhenImp}.

%
%
%
%
%
%
%
%

%
%



{\textit{The drag electric currents.---}}
The drag current of normal electrons as a nonlinear response of the system to the external EM perturbation in the case of longitudinal EM waves reads~\cite{Ivchenkodrag} (see Supplemental Material~\cite{SM} for the details of derivations)
\begin{eqnarray}\label{Eqq11}
\mathbf{j}^{(e)}=
\frac{\mathbf{k}}{2e\omega n}\left|\frac{\sigma_\omega^DE_0}{\varepsilon(\mathbf{k},\omega)}\right|^2,\,\textrm{where}\,\,\,\sigma_\omega^D=\frac{\sigma_0^D}{1-i\omega\tau}.
\end{eqnarray}
The presence of the function $\varepsilon(\mathbf{k},\omega)$ in the denominator here reflects the screening of the external field by the carriers of charge.
It should be noted, that in the presence of the SFs in the system, the drag current of normal electrons is affected by them at plasmon frequencies via their contribution to the dielectric function $\varepsilon(\mathbf{k},\omega)$, as becomes evident from Eq.~(\ref{Eqq5}).

\textcolor{black}{To derive the drag current of fluctuating Cooper pairs, we use the Boltzmann equation~\cite{LarkinVarlamov2005}
\begin{gather}\label{Eqq12}
\dot{f}+\textbf{u}\cdot\partial_\textbf{r} f+2e\Bigl[\textbf{E}(\textbf{r},t)+\textbf{E}^i(\textbf{r},t)\Bigr]\cdot\partial_\textbf{p} f={\cal I}\{f\},
\end{gather}
where $f$ is a distribution function of SFs, $\mathbf{E}^i$ is the induced electric field~\cite{Remark2}, ${\cal I}\{f\}=-(f-\langle f\rangle)/\tau_\textbf{p}$ with $\langle f\rangle$ the locally equilibrium distribution function.
%
We assume that the external EM field causes small perturbation over the homogeneous case, and thus we can expand $f$ and the normal electron density $N$ in powers of external field~\cite{Kittel, Abrikosov}}: $f=f_0+f_1+f_2+o(f_3)$, $N=n+n_1+n_2+o(n_3)$, and $\langle f\rangle=f_0+\partial_n f_0(n_1+n_2)+\partial^2_{n^2}f_0(n_1+n_2)^2/2$.
The latter expansion holds since the equilibrium distribution of fluctuating Cooper pairs depends on the density of normal electrons, as it has been mentioned above, after Eq.~(\ref{Eqq1}).
Furthermore, due to the dependence of the Cooper pairs lifetime $\tau_\textbf{p}$ on normal electron density, it also expands as $\tau_\textbf{p}^{-1}+\partial_n\tau_\textbf{p}^{-1}(n_1+n_2+o(n_3))$.

Decomposing the first-order corrections as plane waves, $f_1(\textbf{r},t)=[f_1\exp(ikx-i\omega t)+f_1^*\exp(-ikx+i\omega t)]/2,\,n_1(\textbf{r},t)=[n_1\exp(ikx-i\omega t)+n_1^*\exp(-ikx+i\omega t)]/2$, and combining all the first-order terms in Eq.~\eqref{Eqq12}, we find 
%
\begin{gather}\label{Eqq13}
f_1=\frac{-2e\tau_\bold{p}\textbf{E}_0\cdot\partial_\textbf{p}f_0+n_1\partial_nf_0}
{1-i(\omega-\textbf{k}\cdot\mathbf{u})\tau_\bold{p}}.
\end{gather}
Obviously, $f_1$ is determined not only by the direct action of the external EM field (the term $\textbf{E}_0\cdot\partial_\textbf{p}f_0$), but also by the normal electron density fluctuations ($n_1$-containing term).
To find 
$n_1$ 
we use the continuity equation, $n_1=\sigma^D_{k\omega}\textbf{k}\cdot\textbf{E}_0/e\omega$.

Onwards, we consider the second-order terms
in Eq.~(\ref{Eqq12}) and find
\begin{gather}
\nonumber
e\textmd{Re}\,\Bigl[\textbf{E}^*_0\cdot\frac{\partial f_1}{\partial\textbf{p}}\Bigr]=
-\frac{1}{\tau_\textbf{p}}\left(f_2-\overline{n}_2\frac{\partial f_0}{\partial n}-\frac{n_1n_1^*}{2}\frac{\partial^2 f_0}{\partial n^2}\right)-\\
\label{Eqq14}
-\frac{\partial\tau_\textbf{p}^{-1}}{\partial n}\textmd{Re}\,\left(f_1-n_1\frac{\partial f_0}{\partial n}\right)\frac{n_1^*}{2},
\end{gather}
where the bar sign stands for the time averaging.
\textcolor{black}{This equation defines the stationary part of the second-order correction $f_2$, which determines the drag current}
\begin{gather}\label{Eqq15}
\textcolor{black}{j^{AL}=2e\int\frac{d\textbf{p}}{(2\pi\hbar)^2}u_xf_2.}
\end{gather}
Due to the integration over the angle in this expression (while taking the 2D integral over $d\mathbf{p}$), all the terms in Eq.~(\ref{Eqq14}) containing the derivative(s) of $f_0$ over $n$ do not contribute to the current~\eqref{Eqq15}.
\begin{figure}
\includegraphics[width=.49\textwidth]{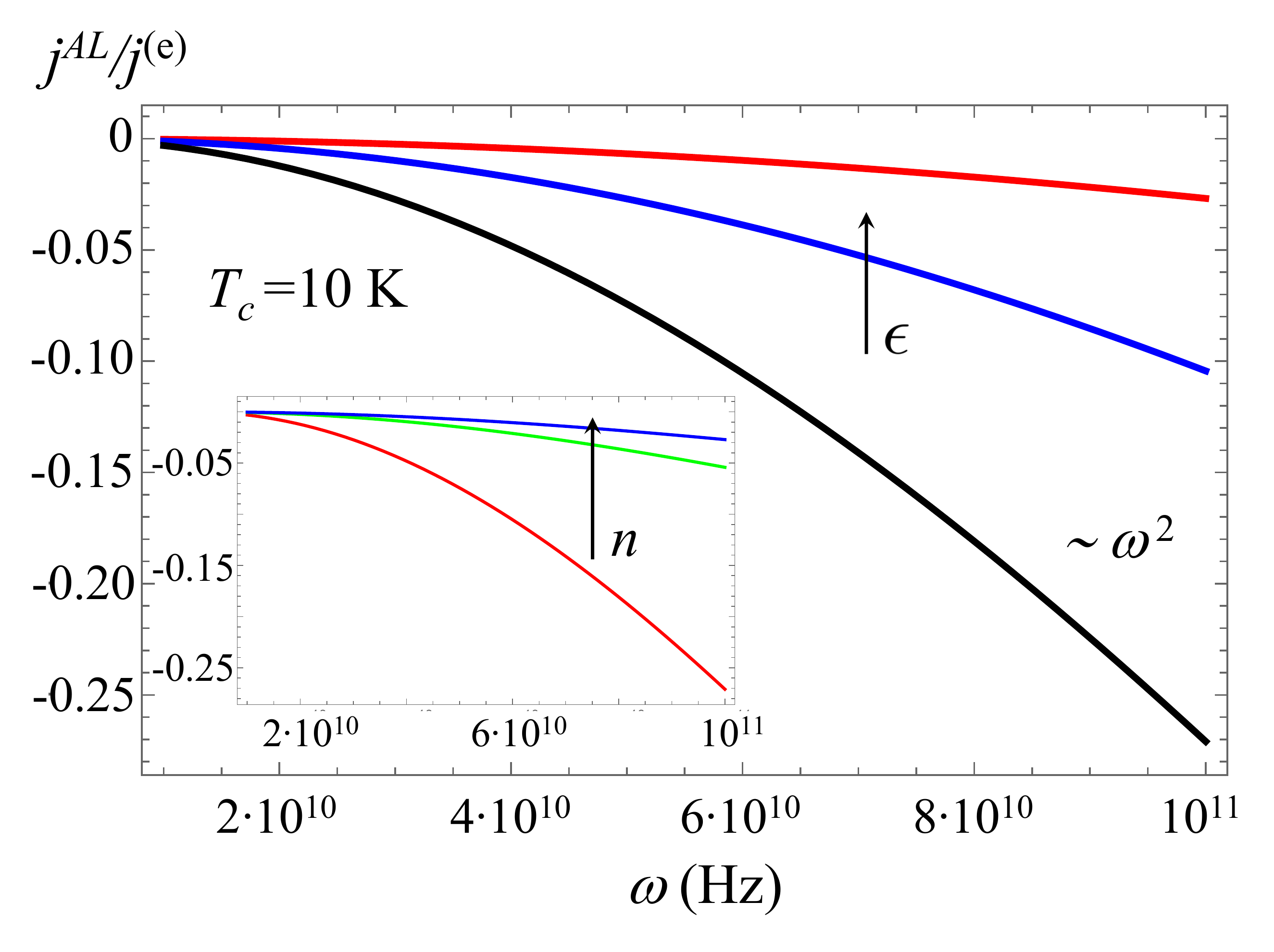}
\caption{
\textcolor{black}{
The ratio of the AL and Drude electric currents~\eqref{Eqq19} as a function of frequency of the external EM field for different temperatures: $\epsilon=(T-T_c)/T_c=0.1$ (red), 0.05 (blue), and 0.03 (black). 
We used $m=0.5~m_0$, where $m_0$ is free electron mass, $\kappa=12$,  $\tau=10^{-9}$~s, and $n=10^{11}~$cm$^{-2}$.
Inset shows the current ratio~\eqref{Eqq19} as a function of frequency for different electron densities: $10^{11}$ (red), $5\cdot10^{11}$ (green), and $10^{12}$~cm$^{-2}$ (blue) for $T=10.3$~K.
}
}
\label{Fig2}
\end{figure}
The remaining terms give the final expression for the second-order correction to the distribution function,
\begin{gather}\label{Eqq16}
f_2=-e\tau_\textbf{p}\textmd{Re}\,\Bigl[\textbf{E}_0^*\cdot\frac{\partial f_1}{\partial\textbf{p}}\Bigr]
-\frac{\tau_\textbf{p}}{2}\frac{\partial\tau_\textbf{p}^{-1}}{\partial n}\textmd{Re}\,\left(f_1n_1^*\right).
\end{gather}
Using 
Eqs.~(\ref{Eqq13}) 
and~(\ref{Eqq14}) 
%
%
and restoring $\varepsilon(\textbf{k},\omega)$ 
we find~\cite{SM}
\begin{gather}\label{Eqq18}
\mathbf{j}^{AL}=
\frac{\mathbf{k}}{2e\omega n}\frac{\sigma_0^{AL}}{\sigma_0^D}\left|\frac{\sigma_\omega^DE_0}{\varepsilon(\mathbf{k},\omega)}\right|^2G\left(\beta_\omega\right),
\end{gather}
where $\beta_\omega=\pi\hbar\omega/16T_c\epsilon$ and
\begin{gather}
\nonumber
G(\beta_\omega)=\frac{1}{\beta_\omega^3}\Bigl\{2\beta_\omega[\beta_\omega\omega\tau-(\beta_\omega+2\omega\tau)\arctan(\beta_\omega)]\\
\label{Eqq18.1}
+(2\beta_\omega-\beta_\omega^2\omega\tau+2\omega\tau)\ln(1+\beta_\omega^2)\Bigr\}.
\end{gather}
Formulas~\eqref{Eqq18}-\eqref{Eqq18.1} represent the second central result of this Letter.


{\textit{Results and discussion.---}}
We can compare the magnitude of the SFs drag current~\eqref{Eqq18} with Eq.~\eqref{Eqq11} describing the drag current of normal electrons,
\begin{gather}\label{Eqq19}
\frac{j^{AL}}{j^{(e)}}=\frac{\sigma_0^{AL}}{\sigma_0^D}G\left(\beta_\omega\right).
\end{gather}
\textcolor{black}{
Figure~\ref{Fig2} shows the spectrum of this ratio.
With the decrease of $\epsilon$ and $n$, the AL correction growth and becomes significant.
}
In the vicinity of the plasmon resonance $\omega=\omega_p$ and at $\omega_p\tau\gg1$, the ratio in Eq.~(\ref{Eqq19}) depends on the value of $\beta_{\omega_p}$.
\textcolor{black}{In the experimentally achievable limit $\beta_{\omega_p}\ll1$~\cite{CommentBigBeta}, we can expand  
$G(\beta_\omega)$ over small $\beta$ and find}
\begin{gather}\label{Eqq20}
\frac{j^{AL}}{j^{(e)}}=-\frac{2}{3}\frac{\sigma_0^{AL}}{\sigma_0^D}\omega_p\tau\beta_{\omega}.
\end{gather}
At the plasmon frequency $\omega=\omega_p$, 
\begin{gather}\label{Eqq21}
\frac{j^{AL}}{j^{(e)}}=-\frac{\pi^2}{96}\frac{e^2k}{(\kappa+1)T_c\epsilon^2}.
\end{gather}
%
%
%
\textcolor{black}{We see, that the dependence of the AL drag current on temperature has a strong singularity $\epsilon^{-2}$ at $T\rightarrow T_c$.}

In Eq.~(\ref{Eqq20}), the smallness of $\beta_\omega$ can be compensated by the large parameter $\omega_p\tau\gg1$ in the vicinity of plasmon resonance, resulting in an experimentally measurable value of SFs drag current.
\textcolor{black}{
Indeed, at $n\sim 10^{11}$~cm$^{-2}$, $k\sim10^2$~cm$^{-1}$~\cite{Kukushkin},  $\omega_p\sim5\cdot10^{10}$~s$^{-1}$.
At the same time, the electron density $n\sim 10^{14}$~cm$^{-2}$ has been recently created in MoS$_2$ material to study the superconducting fluctuations~\cite{MoS2}.
Since $\omega_p\propto\sqrt{n}$, we estimate $\omega_p\sim\,10^{11}$~s$^{-1}$.
Thus, at $\epsilon=0.1$, we find $\beta_{\omega_p}\sim(0.01\div0.2)$.
Taking $\omega_p\tau\sim10$, we estimate the drag current $j^{AL}/j^{(e)}\sim(0.1\div1)\sigma^{AL}_0/\sigma^D_0$.
}

The AL correction gives an increase of conductivity when the system approaches $T_c$. In contrast, the AL correction to the drag effect has negative sign (see Fig.~\ref{Fig2}), as it follows from Eq.~(\ref{Eqq20}).
If the drag current of normal electrons is given by Eq.~(\ref{Eqq11}), SFs give a decrease of the total drag current of the system in the vicinity of $T_c$.
However, if we account for the dependence of the electron relaxation time on its energy, the drag current~(\ref{Eqq11}) might also have negative sign or even change it with frequency~\cite{GlazovReview}.
In this case, the SFs can increase the overall magnitude of the total drag current.

An important and essential feature of Eq.~(\ref{Eqq20}) is that the effect is stronger at bigger $\omega_p\tau$.
It makes us envisage that from the experimental point of view, the photon and acoustic drag effects seem not the best candidates to observe the plasmon amplification of SFs drag current.
Indeed, the acoustic frequencies are much smaller than $\omega_p$, whereas in the photon drag effect the in-plane projection of the photon wave vector is too small to excite  plasmons.
Thus, probably, the most prominent configuration can be the ratchet~\cite{Ivchenko1, Ivchenko2}, when an asymmetric grating structure is deposited above the 2DEG.
Lately, it has been reported that the ratchet current of normal electrons is enhanced at plasmon frequencies~\cite{Kachorovskii}.
Therefore, our calculations suggest the plasmon enhancement of SFs in such structures.

\textcolor{black}{
In recent years, there has emerged a growing interest in terahertz (THz) equilibrium and nonequilibrium studies of different low-dimensional materials in SC regime $T<T_c$~\cite{BeckPRL}. 
It turns out that the THz spectroscopy methods can be utilized to manipulate the SC gap efficiently since they are susceptible. 
In this Letter, we have shown that external EM fields of the THz frequency (which we used in our calculations) can also be used to monitor superconductors in the fluctuating regime.
}

{\textit{Conclusions.---}}
We have considered a two-dimensional material in the vicinity of the transition temperature to a superconducting state, where the superconducting fluctuations can be described by the Aslamazov-Larkin approach~\cite{Comment1Dsystems}.
Using the Boltzmann transport equations, we have studied the dynamics of fluctuations, taking into account the interaction between the Cooper pairs and the normal electron gas within the mean-field random phase approximation approach, and analysed the plasmon resonance phenomenon, showing that it experiences an anomalously large broadening and renormalization of plasmon dispersion caused by the presence of fluctuations in the system~\cite{CommentOutlook}.
This broadening 
has strong sensitivity to  temperature, and it substantially increases when the temperature approaches $T_c$.
Furthermore, we have studied the drag effect of fluctuating Cooper pairs and shown that the drag electric current magnitude is measurable in an experiment.
Our findings open a way for the plasmon spectroscopy (a well-established experimental technique) to serve as an effective tool to test fluctuating phenomena and thus optically explore the properties of superconductors.

We thank A.~Varlamov for fruitful discussions and critical reading of the manuscript.
We acknowledge the support by the Russian Foundation for Basic Research (Project No.~18-29-20033), 
the Ministry of Science and Higher Education of the Russian Federation (Project FSUN-2020-0004), and
the Institute for Basic Science in Korea (Project No.~IBS-R024-D1).




%
%
%
\begin{widetext}

\section{Supplemental Material}

In this Supplemental Material, we provide the details of derivations of the Aslamazov-Larkin (AL) corrections. 
We find (i) the drag current of normal electrons in the presence of superconducting fluctuations (SFs) and (ii) the drag current of SFs themselves.
We also calculate the Landau damping of the plasmons.


\subsection{1. Drag current of normal electrons}
Here we derive the expressions describing the linear and second-order responses of normal (non-superconducting) degenerate electron gas.
We will use the Boltzmann transport equation~\cite{Chaplik},
\begin{gather}\label{App1}
\partial_t\mathcal{F}+\dot{\textbf{p}}\cdot\partial_{\textbf{p}}\mathcal{F}+\dot{\textbf{r}}\cdot\partial_{\textbf{r}}\mathcal{F}=I\{\mathcal{F}\},
\end{gather}
where $\mathcal{F}$ is the distribution function of normal electrons and $I\{\mathcal{F}\}$ is the collision integral, for which we use the model of single--$\tau$ approximation~\cite{Kittel}, which means that $\tau$ is energy-independent and $I\{\mathcal{F}\}=-(\mathcal{F}-\langle \mathcal{F}\rangle)/\tau$.
Here $\langle \mathcal{F}\rangle$ is a locally-equilibrium Fermi-Dirac electron distribution function, which depends on electron density  $N(\textbf{r},t)$ via the chemical potential $\zeta=\zeta(N)$.
We can expand the electron density in series: $N(\textbf{r},t)=n+n_1(\textbf{r},t)+n_2(\textbf{r},t)+o(n_3)$, where $n$ is the equilibrium electron density and $n_i$ are the corrections to the electron density due to external EM field perturbation.

The first nonzero correction to the electric drag current should be found as the second-order response to the external EM field.
Therefore we expand the distribution function in series: $\mathcal{F}=\mathcal{F}_0+\mathcal{F}_1+\mathcal{F}_2+o(\mathcal{F}_3)$, where $\mathcal{F}_0=(\exp\{[\tilde\varepsilon_\mathbf{p}-\zeta(n)]/T\}+1)^{-1}$ is the equilibrium Fermi-Dirac distribution.
We will also need the expansion of locally-equilibrium function $\langle \mathcal{F}\rangle$ with respect to electron density perturbations $\langle \mathcal{F}\rangle=\mathcal{F}_0+(n_1+n_2+...)\partial_n \mathcal{F}_0+(n_1+n_2+...)^2\partial^2 \mathcal{F}_0/\partial n^2/2$.
The electron density fluctuations create the charges in the system and, as a result, they produce the induced electric field which can be found from the Maxwell's equation
$\textmd{div}\, \textbf{D}^i=4\pi\rho$,
where $\mathbf{D}^i=\kappa(z)\mathbf{E}^i$, $\kappa(z)$ is the dielectric function, and $\rho=e(N({\bf r},t)-n)\delta(z)$ is the charge density.
We find
$\textbf{E}^i=-4\pi ie\textbf{k}(N-n)_{{\bf k},\omega}/[(\kappa+1)k]$,
where $\kappa$ is the dielectric constant of the media (semi-infinite layer of the substrate).

For an EM perturbation with the momentum $\textbf{k}$ and assuming that the phase velocity of the wave significantly exceeds the electron velocity $\omega/k\gg v_F$, for a degenerate electron gas at zero temperature [which means here $\partial\zeta/\partial n=2\pi/m$ and $\int d\mathbf{p}\partial_\zeta \mathcal{F}_0(\mathbf{p})/(2\pi)^2=m/(2\pi)$], we find the electric current density (for the normal electrons):
\begin{eqnarray}\label{App2}
\mathbf{j}^{(e)}=
\frac{e\tau}{2m}\frac{\mathbf{k}\sigma_0^D}{\omega}\frac{|E_0|^2}{|\varepsilon(\mathbf{k},\omega)|^2}\frac{1}{1+\omega^2\tau^2}.
\end{eqnarray}
This expression can also be found from the simple consideration of Newton's equations of motion, as it has been reported in Ref.~\onlinecite{Ivchenko2}.
The expression in work mentioned above differs from our result~(\ref{App2}), first, by the numerical factor since we do not take into account the electron spin and, second, by the factor $\varepsilon(\mathbf{k},\omega)$ describing the dynamical screening of external EM perturbation.
%
%


\subsection{2. Drag current of superconducting fluctuations}
Performing the algorithm of analytical derivations discussed in the main text and in Sec.~1 above, we find that the drag current of fluctuating Copper pairs consists of four contributions:
\begin{gather}\nonumber
j_1=-4e^3\left|\frac{{E}_0}{\varepsilon(\mathbf{k},\omega)}\right|^2\int \frac{\tau_\textbf{p}d\mathbf{p}}{(2\pi)^2}
\frac{\partial(u_x\tau_\textbf{p})}{\partial p_x}
\frac{\partial f_0}{\partial p_x}
\mathrm{Re}
\frac{1}{1-i(\omega-ku_x)\tau_\textbf{p}},\\
\nonumber
j_2=\frac{2e^2k}{e\omega}\left|\frac{{E}_0}{\varepsilon(\mathbf{k},\omega)}\right|^2\int \frac{d\mathbf{p}}{(2\pi)^2}
\frac{\partial(u_x\tau_\textbf{p})}{\partial p_x}
\frac{\partial f_0}{\partial n}
\mathrm{Re}
\frac{\sigma^D_\omega}{1-i(\omega-ku_x)\tau_\textbf{p}},\\
\nonumber
j_3=\frac{2e^2k}{e\omega}\left|\frac{{E}_0}{\varepsilon(\mathbf{k},\omega)}\right|^2\int \frac{u_x\tau_\textbf{p}^2d\mathbf{p}}{(2\pi)^2}
\frac{\partial(\tau_\textbf{p}^{-1})}{\partial n}
\frac{\partial f_0}{\partial p_x}
\mathrm{Re}
\frac{(\sigma^{D}_\omega)^*}{1-i(\omega-ku_x)\tau_\textbf{p}},\\
\label{AppII1}
j_4=-\frac{ek^2|\sigma^D_\omega|^2}{(e\omega)^2}
\left|\frac{{E}_0}{\varepsilon(\mathbf{k},\omega)}\right|^2\int \frac{u_x\tau_\textbf{p}d\mathbf{p}}{(2\pi)^2}
\frac{\partial(\tau_\textbf{p}^{-1})}{\partial n}
\frac{\partial f_0}{\partial n}
\mathrm{Re}
\frac{1}{1-i(\omega-ku_x)\tau_\textbf{p}}.
\end{gather}
Not all these terms are equivalent in the order of magnitude of the resulting electric current density.
Below we show, that the leading contributions come from $j_2$ and $j_3$ terms.

Let us consider these terms first.
Taking into account the relations
\begin{gather}\label{AppII2}
\frac{\partial f_0}{\partial n}=-\frac{\partial \mu}{\partial n}\frac{\partial f_0}{\partial\varepsilon_p}
,
\,\,\,\,\,\,
\frac{\partial(u_x\tau_{\textbf{p}})}{\partial p_x}=\frac{\tau_{\textbf{p}}}{2m}\left(1-\frac{2m u_x^2}{\varepsilon_p}\right),
\end{gather}
and
\begin{gather}\label{AppII3}
\mathrm{Re}
\frac{\sigma^D_\omega}{1-i\omega\tau_\textbf{p}}=\sigma_0^D\frac{1-\omega^2\tau\tau_{\textbf{p}}}{(1+\omega^2\tau^2)(1+\omega^2\tau^2_{\textbf{p}})},
\end{gather}
we find
\begin{gather}\label{AppII4}
j_{2}=\frac{k}{e\omega n}\frac{\sigma_0^{AL}\sigma_0^D}{(1+\omega^2\tau^2)}
\left|\frac{E_0}{\varepsilon(\mathbf{k},\omega)}\right|^2\int\limits_1^\infty\frac{dx}{x^3}\frac{x-\omega\tau\beta_\omega}{x^2+\beta_\omega^2},
\end{gather}
where $x=\varepsilon_p/\mu$ and we have disregarded the spatial dispersion of this expression, implying $\omega\gg ku_x$.
The derivative $\partial_n\mu=-\mu/n$ follows directly from the relations $\mu=\alpha T_c\epsilon$, $4m\alpha T_c\xi^2/\hbar^2=1$ and Eq.~(1) from the main text, describing the coherence length $\xi$.

To find $j_3$ contribution, we are using the relation
\begin{gather}\label{AppII5}
\frac{\partial\tau^{-1}_{\textbf{p}}}{\partial n}=\frac{16T_c(p\xi)^2}{\pi n \hbar^3}
\end{gather}
and find
\begin{gather}\label{AppII6}
j_{3}=-\frac{k}{e\omega n}\frac{\sigma_0^{AL}\sigma_0^D}{(1+\omega^2\tau^2)}
\left|\frac{E_0}{\varepsilon(\mathbf{k},\omega)}\right|^2\int\limits_1^\infty\frac{(x-1)^2dx}{x^3}\frac{x+\omega\tau\beta_\omega}{x^2+\beta_\omega^2}.
\end{gather}

Let us further consider the remaining $j_1$ and $j_4$ terms in Eq.~\eqref{AppII1}.
If we disregard the terms $ku_x$ in the denominators (like we did for the second and the third terms assuming $\omega\gg ku_x$), these contributions vanish.
In order to get a nonzero result, one has to keep $ku_x$ in the first order.
It introduces the smallness into the expressions for the curent density, thus lowering their values in comparison with the the second and the third contributions to the drag current.

This simple argument is supported by the direct calculations of these terms.
The results (after analytical integrations) read ($\beta_\omega\ll1$)
\begin{gather}\label{AppII8}
j_1/j^{(e)}\sim\frac{T_c}{\zeta}\beta_\omega^4\left(1+\frac{1}{\omega^2\tau^2}\right),\\\nonumber
j_4/j^{(e)}\sim\left(\frac{k}{k_F}\right)^2\frac{1}{\epsilon}.
\end{gather}
Obviously, both of these terms are small due to the factors $T_c/\zeta\ll1$ and $k/k_F\ll1$ and we can neglect them.

In the mean time, it turns out possible to calculate the integrals in Eqs.~(\ref{AppII4}) and~(\ref{AppII6}) analytically.
This integration yields the final result $j^{AL}\equiv j_{2}+j_{3}$ given in the main text [Eqs.~(19)-(20)].


\subsection{3. The Landau damping}
In this section, we will use a commonly used approach to find the plasmon dispersion and its damping~\cite{RefDassarmaPlasmon}.
We can write down the dispersion relation in the form (accounting for the fact, that our 2D layer lies on top of a semi-infinite material with the dielectric constant $\kappa$)
\begin{eqnarray}
\label{EqFirst}
1-\frac{4\pi e^2}{(\kappa+1) k}
\Pi_{\mathbf{k}\omega}
=0,
\end{eqnarray}
where the polarization operator reads (using $\mathbf{v}=\mathbf{p}/m$)
\begin{eqnarray}
\Pi_{\mathbf{k}\omega}&=&
2\sum_\mathbf{p}
\frac{{\cal F}_\mathbf{p}-{\cal F}_{\mathbf{p}+\mathbf{k}}}{\omega+\tilde\varepsilon_\mathbf{p}-\tilde\varepsilon_{\mathbf{p}+\mathbf{k}}+i\delta}
\approx
-2\sum_\mathbf{p}\frac{\partial {\cal F}_\mathbf{p}}{\partial\tilde\varepsilon_\mathbf{p}}
\frac{\mathbf{v}\cdot\mathbf{k}}{\omega-\mathbf{v}\cdot\mathbf{k}+i\delta}
=
-2\sum_\mathbf{p}
\left(-{\cal F}'_0\right)
\frac{\omega-\omega-\mathbf{v}\cdot\mathbf{k}}{\omega-\mathbf{v}\cdot\mathbf{k}+i\delta}\\
\nonumber
&=&
-2\sum_\mathbf{p}
\left(-{\cal F}'_0\right)
\left(
1-
\frac{\omega}{\omega-\mathbf{v}\cdot\mathbf{k}+i\delta}
\right)
=
-2\sum_\mathbf{p}
\left(-{\cal F}'_0\right)
\left(
1-
\frac{\omega}{\omega-\mathbf{v}\cdot\mathbf{k}}
+i\omega\pi\delta(\omega-\mathbf{v}\cdot\mathbf{k})
\right),
\end{eqnarray}
where ${\cal F}'_0=\partial {\cal F}_\mathbf{p}/\partial\tilde\varepsilon_\mathbf{p}$.
We can now replace the sum by the integral
$\sum_{\mathbf{q}}\rightarrow\int d\mathbf{q}/(2\pi\hbar)^2=\int\int qdqd\phi/(2\pi\hbar)^2$ in cylindrical coordinates, where $\phi$ is the angle between the vectors $\mathbf{v}$ and $\mathbf{k}$.
The integration over the angle gives
\begin{eqnarray}
\int_0^{2\pi}\frac{d\phi}{2\pi}
\left(
1-\frac{\omega}{\omega-vk\cos\phi}
+i\omega\pi\delta(\omega-vk\cos\phi)
\right)
=
1-\frac{|\omega|\theta(\omega^2-k^2v^2)}{\sqrt{\omega^2-k^2v^2}}
+i
\frac{\omega\theta(k^2v^2-\omega^2)}{\sqrt{k^2v^2-\omega^2}}.
\end{eqnarray}
Then
\begin{eqnarray}
\label{EqP1}
\Pi_{\mathbf{k}\omega}
=
-2\int_0^\infty
\frac{pdp}{2\pi\hbar^2}
\left(
-{\cal F}'_0
\right)
\left(
1-\frac{|\omega|\theta(\omega^2-k^2v^2)}{\sqrt{\omega^2-k^2v^2}}
+i
\frac{\omega\theta(k^2v^2-\omega^2)}{\sqrt{k^2v^2-\omega^2}}
\right).
\end{eqnarray}
Furthermore, we can put $
-{\cal F}'_0
=\delta(\tilde\varepsilon_\mathbf{p}-\zeta)$ in the real part of Eq.~\eqref{EqP1} and take this integral, but for the imaginary part of Eq.~\eqref{EqP1} a more careful treatment is required.

Eq.~\eqref{EqFirst} after taking the integration over the real part  reads
\begin{eqnarray}
\label{EqDisp2}
1+\frac{2}{ka_B}
\left(
1-\frac{\theta(\omega^2-k^2v^2)}{\sqrt{1-(kv_F/\omega)^2}}
+iQ(\omega)
\right)
=0,
\end{eqnarray}
where $a_B=\hbar^2(\kappa+1)/2me^2$ and
\begin{eqnarray}
Q(\omega)=\frac{\omega\pi}{m}
\int\limits_0^\infty 
\frac{pdp\theta(k^2v^2-\omega^2)}{\sqrt{k^2v^2-\omega^2}}
\left(
-{\cal F}'_0
\right).
\end{eqnarray}
The plasmon can only exist if its dispersion lies above the electron-hole continuum, i.e. $\omega\gg kv_F$ (otherwise, in the dispersion equation, the real part is smaller than the imaginary part).
Then $1-\left[1-(kv_F/\omega)^2\right]^{-1/2}\approx -0.5(kv_F/\omega)^2$ and~\eqref{EqDisp2} transforms into
\begin{eqnarray}
1-\frac{1}{ka_B}
\left(\frac{kv_F}{\omega}\right)^2
+i\frac{2}{ka_B}Q(\omega)
=0
\end{eqnarray}
or
\begin{eqnarray}
\omega^2=\frac{(kv_F)^2}{ka_B}
-i\frac{2\omega^2}{ka_B}Q(\omega).
\end{eqnarray}
We can find an approximate solution of this equation by successive approximations.
For the first step, we can disregard the damping (put the imaginary part to zero) to find  (using $v_F^2=2\pi\hbar^2 n/m^2$, which accounts for the spin degree of freedom in the concentration $n$)
\begin{eqnarray}
\omega=\frac{kv_F}{\sqrt{ka_B}}=\sqrt{\frac{4\pi e^2nk}{(\kappa+1) m}}\equiv \omega_p.
\end{eqnarray}
The successive approximation gives an equation,
\begin{eqnarray}
\omega^2=\omega_p^2-
i\frac{2\omega_p^2}{ka_B}Q(\omega_p)
=
\omega_p^2\left(1-i\frac{2}{ka_B}Q(\omega_p)\right),
\end{eqnarray}
or expanding,
\begin{eqnarray}
\omega=\omega_p
-i\frac{\omega_p}{ka_B}Q(\omega_p)=\omega_p-i\Gamma_L,
\end{eqnarray}
where $\Gamma_L=\omega_pQ(\omega_p)/ka_B$ is the damping, which we want to find. For that, we should calculate the function
\begin{eqnarray}
\label{EqQp}
Q(\omega_p)&=&\frac{\omega_p\pi}{m}
\int\limits_0^\infty 
\frac{pdp\theta(k^2v^2-\omega_p^2)}{\sqrt{k^2v^2-\omega_p^2}}
\left(
-{\cal F}'_0
\right)
=\omega_p\pi m
\int\limits_{\omega_p/k}^\infty 
\frac{vdv}{\sqrt{k^2v^2-\omega_p^2}}
\left(
\frac{1}{T}
\frac{\exp{[(\tilde\varepsilon_{\mathbf{p}}-\zeta)/T]}}{\left(\exp{[(\tilde\varepsilon_{\mathbf{p}}-\zeta)/T]}+1\right)^2}
\right)\\
\nonumber
&=&
\pi
\frac{\omega_p}{|\omega_p|}
\frac{m}{2T}
\frac{\omega_p^2}{k^2}
\int\limits_{1}^\infty 
\frac{dx}{\sqrt{x-1}}
\frac{\exp{[x\left(\frac{\omega_p}{kv_T}\right)^2-\frac{\zeta}{T}]}}{\left(\exp{[x\left(\frac{\omega_p}{kv_T}\right)^2-\frac{\zeta}{T}]}+1\right)^2},
\end{eqnarray}
where we used the substitution $x=k^2v^2/\omega_p^2$ and defined the thermal velocity as $v_T=\sqrt{2T/m}$.

If $\zeta\gg T$, it is equivalent to $v_F\gg v_T$. 
However, as we have discussed before, $\omega_p\gg kv_F$.
Combining these two inequalities, we find $\omega_p\gg kv_T$. It means that the exponential factors in Eq.~\eqref{EqQp} are huge, and we can rewrite (neglecting 1 in the last denominator of Eq.~\eqref{EqQp})
\begin{eqnarray}
Q(\omega_p)
=\pi\textrm{sgn}(\omega_p)
\left(\frac{\omega_p}{kv_T}\right)^2
\int\limits_{1}^\infty 
\frac{dx}{\sqrt{x-1}}
\left(
\exp{\left[-x\left(\frac{\omega_p}{kv_T}\right)^2+\frac{\zeta}{T}\right]}
\right)
=
\pi\textrm{sgn}(\omega_p)
\left(\frac{\omega_p}{kv_T}\right)^2
e^{\zeta/T}
\sqrt{\pi\frac{k^2v^2_T}{\omega_p^2}}e^{-\omega_p^2/k^2v_T^2}.
\end{eqnarray}
Then
\begin{eqnarray}
\Gamma_L=\frac{\omega_p}{ka_B}
\left(\frac{\omega_p}{kv_T}\right)
\pi\sqrt{\pi}
e^{\zeta/T-\omega_p^2/k^2v_T^2}
=
\pi\sqrt{\pi}
\frac{v_T}{a_B}
\left(\frac{\omega_p}{kv_T}\right)^2
\exp{\left[-\frac{\omega_p^2/k^2-v_F^2}{v_T^2}\right]}.
\end{eqnarray}
%
Evidently, $\Gamma_L$ is exponentially small within the range, where the plasmons exist, $\omega_p\gg kv_F\gg kv_T$ at $\zeta/T\gg 1$.

%
%
%
\end{widetext}
%
%
%



\begin{thebibliography}{50}
\bibitem{CommentBoseMixture}
Thus, we consider the system as a mixture of two interacting gases, one is the degenerate gas of normal-state electrons and the other is the Bose gas of fluctuating Cooper pairs.
\bibitem{CommentGauss} We will be using the cgs (Gauss) units.
\bibitem{CommentRegPlasmon}
In the absence of SFs ($\sigma^{AL}_0=0$), Eq.~\eqref{Eq9} gives the conventional expression for the plasmon frequency of an electron gas~\cite{Volkov}, 
%
\begin{gather}
\nonumber
\omega=\omega_p\sqrt{1-\left(\frac{1}{2\omega_p\tau}\right)^2}
-\frac{i}{2\tau}.
\end{gather}
%
Evidently, the plasmon branch can only exist if the scattering is weak, $2\omega_p\tau>1$. 
In actual experiments, a pronounced resonance is observable under the condition $\omega_p\tau\gg1$~\cite{Kukushkin}.
\bibitem{CommentBigBeta}
\textcolor{black}{Eq.~(\ref{Eq11}) is derived under the condition $\beta_{\omega_p}\ll1$. 
Simple estimations show, that the opposite case $\beta_{\omega_p}\gg1$ seems hardly realizable in experiments. 
Indeed, to have  $\beta_\omega\sim10$, we need $\omega_p\sim 10^{13}$~s$^{-1}$, which corresponds to 2D electron gas density $n\sim 10^{18}$~cm$^{-2}$, which is usually too high.}
\bibitem{CommentWhenImp}
Let us  estimate the temperatures, at which 
$\Gamma_i$ surpasses $\Gamma_s=\eta\omega_p$. 
The impurities are dominant  when 
$\Gamma_i>\eta\omega_p$, which gives $\sigma_0^D/\sigma_0^{AL}>(\omega_p\tau)^2$ or $\epsilon>2\pi(\omega_p\tau)^2/(16k_Fl)$, where $k_F$ is the electron Fermi wave vector and $l$ is the mean free path. 
Thus, for $\omega_p\tau\sim10$ and $k_Fl\sim10$, we find $\epsilon\geqslant 2$. Thus if $T_c\sim10$~K, we estimate $T>3T_c=30$~K. 
%
%
%
At finite temperatures, there emerges another contribution originating from the Landau damping of plasmon oscillations of normal electrons~\cite{LandauDamping, Raether}.
Its magnitude for 2D electrons reads~\cite{SM} 
%
\begin{eqnarray}
\nonumber
\Gamma_{L}
=
\frac{2e^2\pi^{3/2}}{\hbar^2}
\frac{mv_T}{\kappa+1}
\left(\frac{\omega_p}{kv_T}\right)^2
\exp{\left[-\frac{\omega_p^2/k^2-v_F^2}{v_T^2}\right]},
\end{eqnarray}
%
where $v_T=\sqrt{2T/m}$.
Due to $\omega_p/k\gg v_F\gg v_T$, we conclude that the Landau damping has a negligibly small value at temperatures 
$T\sim T_c$ since $T_c/\zeta\ll1$, where $\zeta$ is the Fermi energy of electrons in the normal state.
Indeed, we can easily estimate that for $\omega_p=10^{11}$~Hz and $k=10^2$~cm$^{-1}$, $\omega_p/kv_T\sim10^3$ for $T\sim1000$~K.
\bibitem{Comment1Dsystems}
We want to mention that our formalism allows considering one-dimensional systems, where the AL conductivity scales as $\epsilon^{-3/2}$~\cite{LarkinVarlamov}, as compared with the 2D case, where it is $\sim\epsilon^{-1}$, but we leave this issue beyond the scope of this Letter.
\bibitem{CommentOutlook}
We have considered one particular type of fluctuating effects: the Aslamazov-Larkin correction.
In certain situations, other contributions, in particular, the Maki-Thompson~\cite{Maki,Thomson} and the ``density of states''~\cite{ALDOS} can play a significant role.
The Boltzmann equations approach cannot be adapted to these corrections, and to study them a microscopic examination is required, which is beyond the scope of this Letter.
Nevertheless, the dispersion relation describing the combined action of normal electrons and SFs given in Eq.~(\ref{Eqq5}) has a general form independent of the type of contribution to conductivity caused by the SFs.
The other corrections will just appear as additional terms in $\sigma^{AL}_{k\omega}$.
In the mean time, this simple modification of the theory is not applicable to the calculations of the drag current.
There one is bound to utilize the quantum approaches instead of the semiclassical Boltzmann equations.
\end{thebibliography}
\end{document}